\documentclass[runningheads]{llncs}
\usepackage{graphicx}
\usepackage{amsmath,amssymb} % define this before the line numbering.
\usepackage{color}
\begin{document}

\mainmatter
\def\ACCV20SubNumber{157}  % Insert your submission number here

%===========================================================
\title{FREA-Unet: Frequency-aware U-net for Modality Transfer}

\author{Hajar Emami, Qiong Liu, Ming Dong}
\institute{hajar.emami.gohari@wayne.edu, qiongliu1@gmail.com, mdong@wayne.edu}

\maketitle

%===========================================================
\begin{abstract}
While Positron emission tomography (PET) imaging has been widely used in diagnosis of number of diseases, it has costly acquisition process which involves radiation exposure to patients. However, magnetic resonance imaging (MRI) is a safer imaging modality that does not involve patient's exposure to radiation. Therefore, a need exists for an efficient and automated PET image generation from MRI data. In this paper, we propose a new frequency-aware attention U-net for generating synthetic PET images. Specifically, we incorporate attention mechanism into different U-net layers responsible for estimating low/high frequency scales of the image. Our frequency-aware attention U-net computes the attention scores for feature maps in low/high frequency layers and use it to help the model focus more on the most important regions, leading to more realistic output images. Experimental results on 30 subjects from Alzheimers Disease Neuroimaging Initiative (ADNI) dataset demonstrate good performance of the proposed model in PET image synthesis that achieved superior performance, both qualitative and quantitative, over current state-of-the-arts.
%\dots
\end{abstract}

%===========================================================
\section{Introduction}
Positron emission tomography (PET), a nuclear medical imaging technology offers potential for diagnosing a number of neurological diseases such as Alzheimer, Epilepsy, Head and Neck Cancer by reflecting tissue metabolic activity in brain. However, obtaining high‐quality PET images is costly and requires radioactive substance injection into the body, which may cause side effects to patients. 
Due to these reasons, baseline datasets usually have smaller number of PET cases compare to other modalities. For example, in Alzheimer’s Disease Neuroimaging Initiative (ADNI) database, only approximately half of subjects have PET scans. 
On the other hand, magnetic resonance imaging (MRI) has excellent soft tissue contrast with anatomical details that does not involve patient's exposure to radiation. 
In order to circumvent the problems of PET imaging acquisition and streamline clinical efficiency, recent attention has been given to estimate PET image from its corresponding MRI data. Eliminating PET acquisition reduces time, costs and side effects to patients. 

\textbf{Related work.}
To date, a number of computational methods have been proposed for medical image synthesis using machine learning approaches. Several researches employed conventional machine learning methods, such as Gaussian Mixture Model (GMM), structure random forest (SRF), etc to generate missing type of medical image modality from available image modalities. In \cite{kang2015prediction}, the authors employed a regression forest for predicting the brain PET images using MRI inputs. Huynh et al. \cite{huynh2015estimating} used structured random forest together with an auto-context model to estimate CT patches from their corresponding MRI patches. At the end, all generated CT patches from a given MRI input are combined together to generate the corresponding CT image.

Recently, many deep learning models have been successfully applied to learning the mappings between two different image domains \cite{choi2018stargan,huang2018multimodal,isola2017image,liu2017unsupervised,murez2018image,zhu2017unpaired,zhu2017toward}. These deep learning approaches have been employed in medical imaging for generating missing type of modality from available type of imaging modality, e.g., generating synthetic CT images from MR images \cite{nie2016estimating,han2017mr,emami2018generating,emami2020attentionArxive} or estimating PET images from MRI data \cite{li2014deep,pan2018synthesizing,armanious2020medgan}. 

Deep convolutional neural networks (CNNs), a type of multi-layer deep learning models can be used to capture nonlinear mappings between input and output image domains. The CNN models are trained by minimizing the voxel-wise differences with respect to the generated output images and the ground truth data. 
CNN models were employed to perform image-to-image mapping between MRI and CT images in \cite{nie2016estimating}. Li \cite{li2014deep} developed a deep learning approach in which 3D-CNN is employed to estimate the output PET images given the input MRI modality. 
A U-net architecture, a well-established fully convolutional network (FCN) first introduced in \cite{ronneberger2015u} for biomedical image segmentation. 
In \cite{han2017mr}, Han developed a learning-based approach in which U-net architecture is employed to generate synthetic CT images from T1-weighted MR data. 

Generative adversarial networks (GANs) \cite{goodfellow2014generative}, another type of deep learning models with two competing networks: a generator to synthesize output images and a discriminator to distinguish between the synthesized and real data have been used in image generation tasks. 
Recently, GANs have been used in medical image-to-image translation tasks \cite{gehari2018generating,armanious2020medgan,pan2018synthesizing,nie2017medical,emami2018generating,emami2020attention}. Pan \cite{pan2018synthesizing} used a GAN model to capture the underlying relationship between MRI and PET images to generate missing PET data. 
In the image-to-image translation task, the models learn the mapping between two image domains using a training set of paired data. Since paired training data are not available in many tasks, Zhu in  \cite{zhu2017unpaired} intoduced CycleGAN to translate an image from a source domain to a target domain in the absence of paired data. Unsupervised image to image translation using cycleGAN was used in \cite{wolterink2017deep} to learn the mapping between unpaired MRI and CT imaging modality. 

Attention plays an important role in human visual process for building a visual representation and possessing context. Inspired from human attention mechanism \cite{rensink2000dynamic}, attention-based deep learning models have been used in a variety of computer vision and machine learning tasks including image classification \cite{xiao2015application,zhou2016learning}, image segmentation \cite{chen2016attention}, image-to-image translation \cite{mejjati2018unsupervised,chen2018attention,emami2020spa}, natural language processing \cite{eriguchi2016tree,lin2017structured} and time series forecasting \cite{siridhipakul2019multi,aliabadi2020attention}. 
Attention can be defined as a scalar matrix showing the relative importance of activation maps at different spatial locations \cite{simonyan2013deep}. Incorporating attention into deep learning models helps to improves the performance by focusing on the most relevant features.
Zhou et al. \cite{zhou2016learning} have shown that attention maps can be used to localize the object of interest and improve object localization accuracy. 
In \cite{zhou2016learning}, the attention in classification task is produced by removing top average-pooling layer of a classifier CNN and helped to identify the discriminative features across classes. 
Zagoruyko et al. \cite{zagoruyko2016paying} improved the performance of a small student CNN network by transferring the attention knowledge from a larger and more powerful teacher CNN. They defined the attention based on the assumption that the absolute value of a neuron activation is relative to the importance of that neuron in order to classify the input image in classification task. 

Jetley et al. \cite{jetley2018learn} incorporated attention mechanism into a classifier CNN to amplify the relevant features and suppress the irrelevant features for classification of the input. Their proposed approach bootstrapped standard CNN architectures in classification. 
Recent studies show that incorporation of attention learning in both image generation and image-to-image translation tasks leads to more realistic output. Zhang et al. \cite{zhang2018self} proposed self-attention GAN that uses a self-attention mechanism in image generation task. Chen et al. and Mejjati et al. \cite{chen2018attention,mejjati2018unsupervised} used an attention network to localize the object of interest and excluding the background in image-to-image translation task to improve the quality of generated images.  

\begin{figure}[t]
\centering
\includegraphics[width=0.5\linewidth]{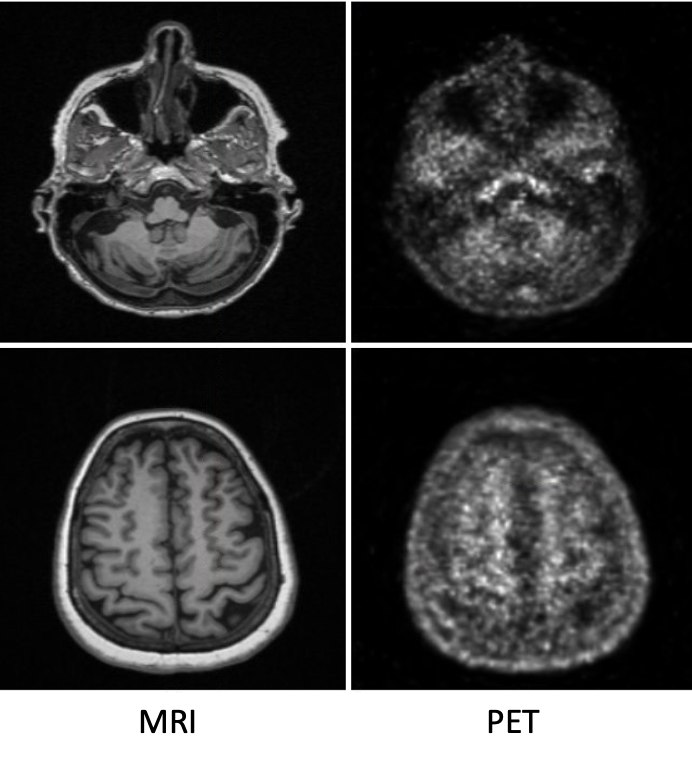} 
\caption{
Examples of brain MRI datasets (the first column) with the corresponding PET images of the same patient (the second column), highlighting significant difference in MRI and PET texture.}
\label{fig:Examples}
\end{figure}

\textbf{This work.}
While deep learning models are great potential in medical imaging, generating synthetic PET images from MRI data is a challenging task due to their different appearances. In Fig.~\ref{fig:Examples}, two examples of brain MR images (the first column) are shown with the corresponding PET images of the same patient (the second column) with a clear different textures. While MR images show richer texture information than PET images, PET images contain complicated texture with different frequency scales. This would impact synthetic PET generation, suggesting that separately optimizing on different frequency scales can lead to more realistic output with more preserved details in PET image generation. 

In this paper, we propose a novel deep learning method, a frequency-aware U-net model (FREA-Unet), that generates low/high frequency scales of the image separately in different layers for estimating synthetic PET images. Optimizing low/high frequency scales of the image with separate loss functions can help to balance between the error of the predicted PET image and the image sharpness and resolution.

We also use end-to-end-trainable attention modules in our FREA-Unet to weight different features in low/high frequency scales. 
Specifically, the attention in the low/high frequency layers is defined as the compatibility scores \cite{jetley2018learn} between feature maps extracted at frequency layers and the output from the second to the last layer in U-net's decoding part showing the relevant features for generating the output image. 
The extracted spatial attention scores are used in the scale layers so that higher weights are given to the relevant regions, leading to more realistic output images. 
The major contribution of our work is summarized as follows: 
\begin{itemize}
  \item We proposed a novel frequency-aware U-net model for medical imaging modality transfer, FREA-Unet. Based on the proposed model, we use a modified loss function with different weights to optimize different scales during FREA-Unet training.
  \item We also incorporated attention mechanism into low/high frequency layers to focus more on the relevant features during image translation.
  \item We proposed a complete end-to-end PET image generation model from MRI data. 
  \item FREA-Unet demonstrates the effectiveness of separating low/high frequency generation and also incorporating the attention mechanism into the image generation model. Through extensive experiments, we show that, both qualitatively and quantitatively, FREA-Unet significantly outperforms other state-of-the-art methods on ADNI dataset, offering potential PET/MRI applications.
\end{itemize}

\section{Materials and methods}
\subsection{FREA-Unet}
\begin{figure}[t]
\centering
\includegraphics[width=1\linewidth]{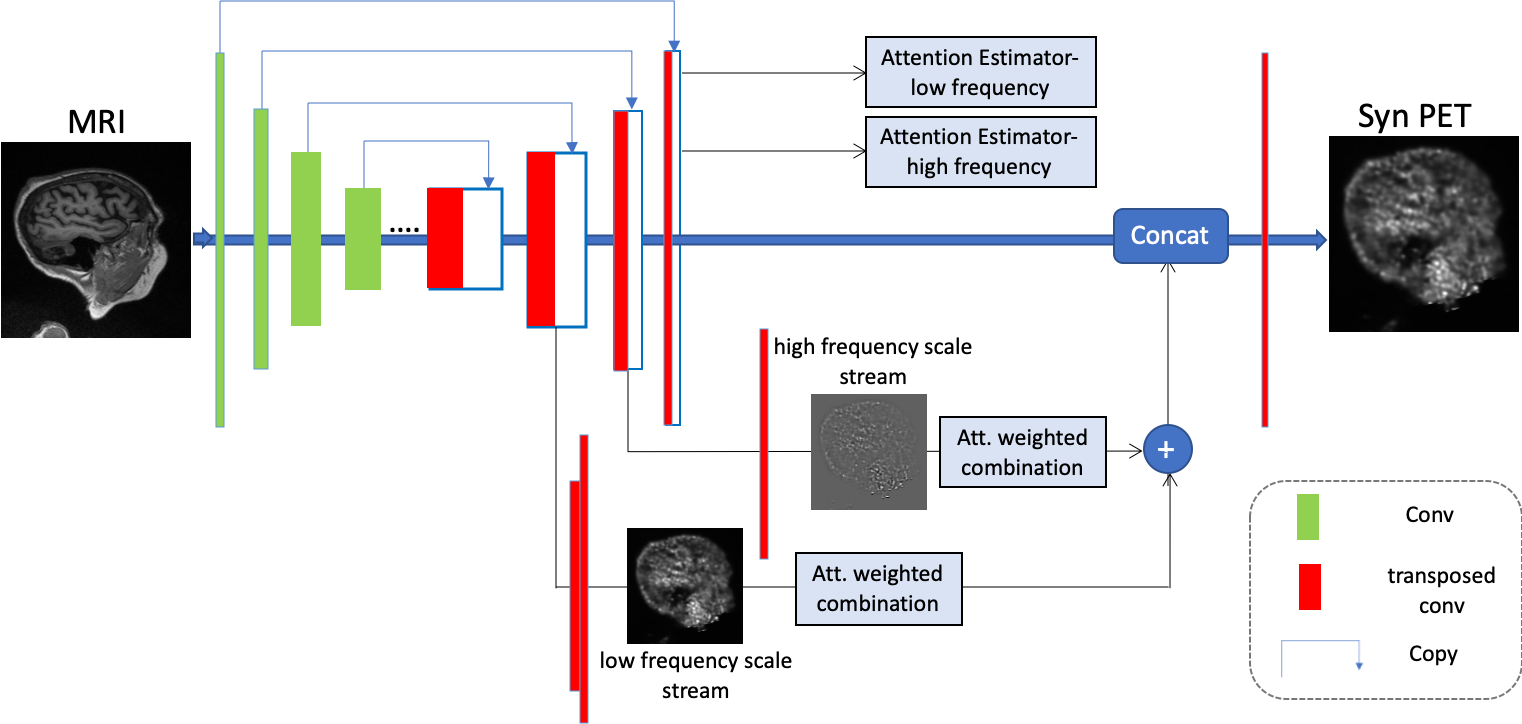} 
\caption{
FREA-Unet architecture. Low and high frequency PET scales are generated in two different layers of FREA-Unet's decoding path. Each frequency scale generation stream has one attention module to weight different features. The outputs of two frequency scale streams are fused and fed into last layer of the decoder to generate the final output PET image.}
\label{fig:arch}
\end{figure}

The goal of the proposed FREA-Unet model is to estimate a mapping $F_{MR \rightarrow PET}$ from the source image domain (MRI) to the target image domain (PET). The mapping F is learned from paired training data $S = \{(mr_i,pet_i)|mr_i \in MR, pet_i \in PET, i=1,2,...,N\}$, where N is the number of MRI-PET image pairs. 
Generating low/high frequency scales of the image separately with different loss functions can help the model to generate more realistic output with more preserved low/high frequency details.
The proposed FREA-Unet model achieves this by explicitly generating low and high frequency PET scales in two different layers of the U-net's decoder as pictured in Fig.~\ref{fig:arch}. 

There are two end-to-end-trainable attention modules for low/high frequency layers to weight different features based on their importance in PET image generation. Incorporating the attention mechanism into the layers responsible for generating low/high frequency scales helps the FREA-Unet model to focus more on the more relevant features during image translation leading to more realistic output images. 

\subsection{Frequency-aware network}
U-net architecture has two symmetrical paths: an encoding path and a decoding path to leverage both local and hierarchical information in order to estimate more accurate reconstruction of the encoded input. Similar to a regular CNN, the encoding path of the U-net contains convolutional layers to encode the context of the input image. On the other hand, the decoding path includes transposed convolutional layers to reconstruct the estimated image.
The U-Net architecture has direct connections across its contracting path and expanding path. These skip connections combine high resolution features from the contracting path with the up-sampled features from the expanding path and use them as the inputs to the convolutional layers in the expanding path. This typically leads to improved resolution for the network output. 

In our proposed FREA-Unet’s contracting path, we have six convolutional layers with 64, 128, 256, 512, 512, 512 filters, respectively. Each convolutional layer uses these filters to perform 2D convolutions on its input. The input of each layer is the output of its previous layer, except for the first layer in which the input is MRI image. The convolution outputs are followed by a nonlinear activation function: Rectified Linear Unit (ReLU), and batch normalization. Batch normalization allows using much higher learning rates and accelerates the network training. In the expanding path of FREA-Unet architecture, we have six transposed convolutional layers with 512, 1024, 1024, 512, 256, 128 filters, respectively, followed by ReLU and batch normalization. In addition, we use dropout in layers of the FREA-Unet as an effective technique for regularization and preventing the co-adaptation of neurons in the network. 

In the proposed FREA-Unet architecture, we assign two different layers in the decoding path for low/high frequency image generation in order to generating low/high frequency PET images separately. Specifically, we consider the fourth layer of the decoding path for low frequency generation and the fifth layer of the decoding path for high frequency PET image generation. Low/high frequency scales are separated using a Gaussian filter for each image during the training phase so the model can optimize different scales separately. 
As it is illustrated in Fig.~\ref{fig:arch}, low/high frequency scale outputs are combined after generation. 
As shown in our experimental results in Section 3, optimizing on different scales helps generate more realistic PET images with improvement in quantitative results. 

Learning the end-to-end mapping function from MRI input and low/high frequency scales requires the estimation of network parameters achieved through minimizing the prediction error between the predicted images $F(MR;\theta)$ and the corresponding ground truth PET images.
The objective for the mapping $F_{MR \rightarrow PET_{low}}$ can be defied as:
\begin{equation}
 L_{low}(F) =||F(mr;\theta_{low})-pet_{low}||_2
 \label{equ:Llow}
\end{equation}

Similarly, the objective for updating the mapping $F_{MR \rightarrow PET_{high}}$ in high frequency stream of the generator is defined as:
\begin{equation}
 L_{high}(F) =||F(mr;\theta_{high})-pet_{high}||_1
 \label{equ:Lhigh}
\end{equation}

Since $L_1$ norm leads to less blurry results in image generation tasks \cite{isola2017image}, we use $L_1$ norm in our high scale generation stream. 
Finally, by combining low/high frequency streams loss functions, the full objective of the FREA-Unet model can be expressed as:
\begin{equation}
 L_{FREA-Unet} = L_{low}(F) + L_{high}(F)
 \label{equ:LFREA}
\end{equation}

\subsection{Attention module}
In the proposed FREA-Unet model, low/high frequency scales are optimized separately. We incorporate attention mechanism into low/high scale layers to weight different features based on their importance in the output PET image. Specifically, the attention in the low/high frequency layers is defined as the compatibility scores \cite{jetley2018learn} between feature maps extracted at frequency layers and the output from the second to the last layer in FREA-Unet's decoding path showing the relevant features for generating the output image. 

We define the compatibility score as result of the dot product between the feature map $f_i$ and the output of the second to last layer of the model that has the input image size, having only to pass through the final layer to produce the output PET image $O_{pet}$:
\begin{equation}
  c_i = \langle f_i, O_{pet}  \rangle, i \in \{1....n\}
  \label{equ:dt}
\end{equation}

The extracted spatial attention scores are used in the low/high scale layers so that higher weights are given to the relevant regions, leading to more realistic output images. 

\subsection{Datasets}
The brain dataset was acquired from 30 subjects with both MRI and PET scans in the Alzheimer’s Disease Neuroimaging Initiative (ADNI) database (www. adni-info.org). ADNI provides a large database of studies with the goal of understanding the development and pathology of Alzheimer’s Disease. Subjects are diagnosed as cognitively normal (CN), significant memory concern (SMC), early mild cognitive impairment (EMCI), mild cognitive impairment (MCI), late mild cognitive impairment (LMCI) or having Alzheimer’s Disease (AD). 
All 30 PET subjects obtained from ADNI datasets were aligned to their corresponding T1- weighted MRI using coregistration so there is spatial correspondence MRI and PET slices in each subject. 
We cropped input MRI and PET images to the size of 256$\times$256 for training.

\section{Experimental Results}
\subsection{Experimental Setup}
Extensive experiments were performed to compare FREA-Unet with current sate-of-the-art models for generating synthetic PET images from MRI inputs. Based on the same training and testing splits, the following models used previously for similar image generation tasks are included in our empirical evaluation and comparison.

\begin{itemize}
\item \underline{U-net model}: Following techniques developed by Han \cite{han2017mr}, we implemented and trained U-net model to compare against the FREA-Unet approach. The U-net model used in our empirical evaluation has no attention mechanism and generates only one frequency scale output. The architecture includes six convolutional layers in the encoding path and six transposed convolutional layers in the decoding path.

\item \underline{pix2pix}: We implemented and trained pix2pix model developed by Isola \cite{isola2017image}, with five residual blocks in the generator. The discriminator is a regular CNN with five convolutional layers that classified the input image as real or synthetic. 

\item \underline{cGANs with U-Net architecture (UcGAN)}: We implemented and trained conditional GAN (cGAN) model with U-net \cite{ronneberger2015u} architecture as its generator. We include comparison with UcGAN as FREA-Unet's core architecture is also based on the U-net. The discriminator is a regular CNN with five convolutional layers.
\end{itemize}

\subsection{Training and Implementation Details}
To evaluate the performance, three-fold cross-validation \cite{stone1974cross} was used in our model training and testing. That is, out of the 30 cases, 20 cases are randomly selected for the training, and the remaining cases are used to test the trained model. We repeat the experiment three times and report the averaged outcome.

The weights in FREA-Unet were all initialized from a Gaussian distribution with parameter values of 0 and 0.02 for mean and standard deviation, respectively. The model is trained with ADAM optimizer with an initial learning rate of 0.0002 and with a batch size of 1. We trained the model for 200 epochs on a NVIDIA GTX 1080 Ti GPU. We used the same training setting for all models in our comparison.

\subsection{Quantitative Measurement}
We used three commonly-used quantitative measures to evaluate the prediction accuracy of the models between the generated PET images and the real PET. These include Mean Absolute Error (MAE) defined below:
\begin{equation}
MAE =  \frac{\sum_{i=1}^H |realPET(i) - synPET(i)}{H}
  \label{equ:mae}
\end{equation}
where H is the number of body voxels. The second measure is the Structural Similarity Index (SSIM) defined as:
\begin{equation}
SSIM =  \frac{(2\mu_{PET} \mu_{synPET} + C_1)(2\sigma_{PETsynPET} + C_2)}{(\mu_{PET}^2 + \mu_{synPET}^2 + C_1)(\sigma_{PET}^2 + \sigma_{synPET}^2 + C_2)}
  \label{equ:ssim}
\end{equation}
where $\mu_{PET}$ denotes the mean value of PET image, $\mu_{synPET}$ denotes the mean value of generated PET image, $\sigma_{PET}^2$ is the variance of image PET, $\sigma_{synPET}^2$ is the variance of image synPET, and the parameters $C1 = (k1Q)^2$ and $C2 = (k2Q)^2$ are two variables to stabilize the division with weak denominators, where $k1 = 0.01$ and $k2 = 0.02$. The last measure is the Peak Signal-to-Noise Ratio (PSNR) defined as:
\begin{equation}
PSNR = 10log_{10} ( \frac{Q^2}{MSE})
  \label{equ:psnr}
\end{equation}
where $Q$ is the maximal intensity value of PET and synPET images, and MSE is the mean squared error.

For the MAE metric, the lower value means better prediction results, that is, more realistic generated PET images. For both SSIM and PSNR, higher values indicate better prediction.

\subsection{Ablation Study}
We first performed model ablation to evaluate the impact of each component of FREA-Unet.
In Table 1, we report MAE, PSNR and SSIM for different configurations of our model. 
First, we removed the attention mechanism from the model and also optimized the model on the original input scale (without seperating low/high frequency scales). As a consequence, we also used the regular U-net loss. In this case, our model is reduced to the U-net architecture (U-net) with MAE: 52.56 $\pm$ 6.71, PSNR: 26.75 $\pm$ 1.96 and SSIM: 0.84 $\pm$ 0.05. 

Next, we removed "optimization on different scales" but kept attention mechanism to weight different features (FREA-Unet-wo-Freq). Our results show that this leads to better results (MAE: 51.10 $\pm$ 7.03, PSNR: 27.08 $\pm$ 1.33 and SSIM: 0.85 $\pm$ 0.04) than U-net, but worse than FREA-Unet. Clearly, separately optimizing low/high scales can help to achieve more accurate synPETS with preserved organ structures.

Further, we removed the attention mechanism from our model and optimized the model on low/high scales separately (FREA-Unet-wo-Att). This results to a higher MAE (48.19 $\pm$ 6.42) and lower PSNR (27.93 $\pm$ 1.84) and SSIM (0.86 $\pm$ 0.04) when compared with the full version of FREA-Unet. This ablation study showed that FREA-Unet works better when attention is employed to weight different features based on their importance in estimating PET images. Separately optimizing low/high scales using different weights obtained from attention modules helped to generate more realistic synthetic PETs with higher spatial resolution. 

\vspace{0.5cm}

\setlength{\tabcolsep}{6pt}
\begin{table}[h]
\centering
\vspace{-0.6cm}
\caption{Performance comparison for ablations of FREA-Unet in synPET generation.}
\vspace{-0.3cm}
\label{table:ablation}
\begin{tabular}{lclclclcl}
\hline
$\qquad\qquad$& MAE & PSNR & SSIM\\
\hline
U-net & 52.56 $\pm$ 6.71 & 26.75 $\pm$ 1.96 & 0.84 $\pm$ 0.05\\
FREA-Unet-wo-Freq & 51.10 $\pm$ 7.03 & 27.08 $\pm$ 1.33 & 0.85 $\pm$ 0.04\\
FREA-Unet-wo-Att & 48.19 $\pm$ 6.42 & 27.93 $\pm$ 1.84 & 0.86 $\pm$ 0.04\\
FREA-Unet & {\bf 46.26 $\pm$ 5.03} & {\bf 28.45 $\pm$ 2.01} & {\bf 0.87 $\pm$ 0.04} \\
\hline
\end{tabular}
\vspace{-0.5cm}
\end{table}
\setlength{\tabcolsep}{1.6pt}

\subsection{Performance of Image Synthesis Models}
\begin{figure}[t]
\centering
\includegraphics[width=1\linewidth]{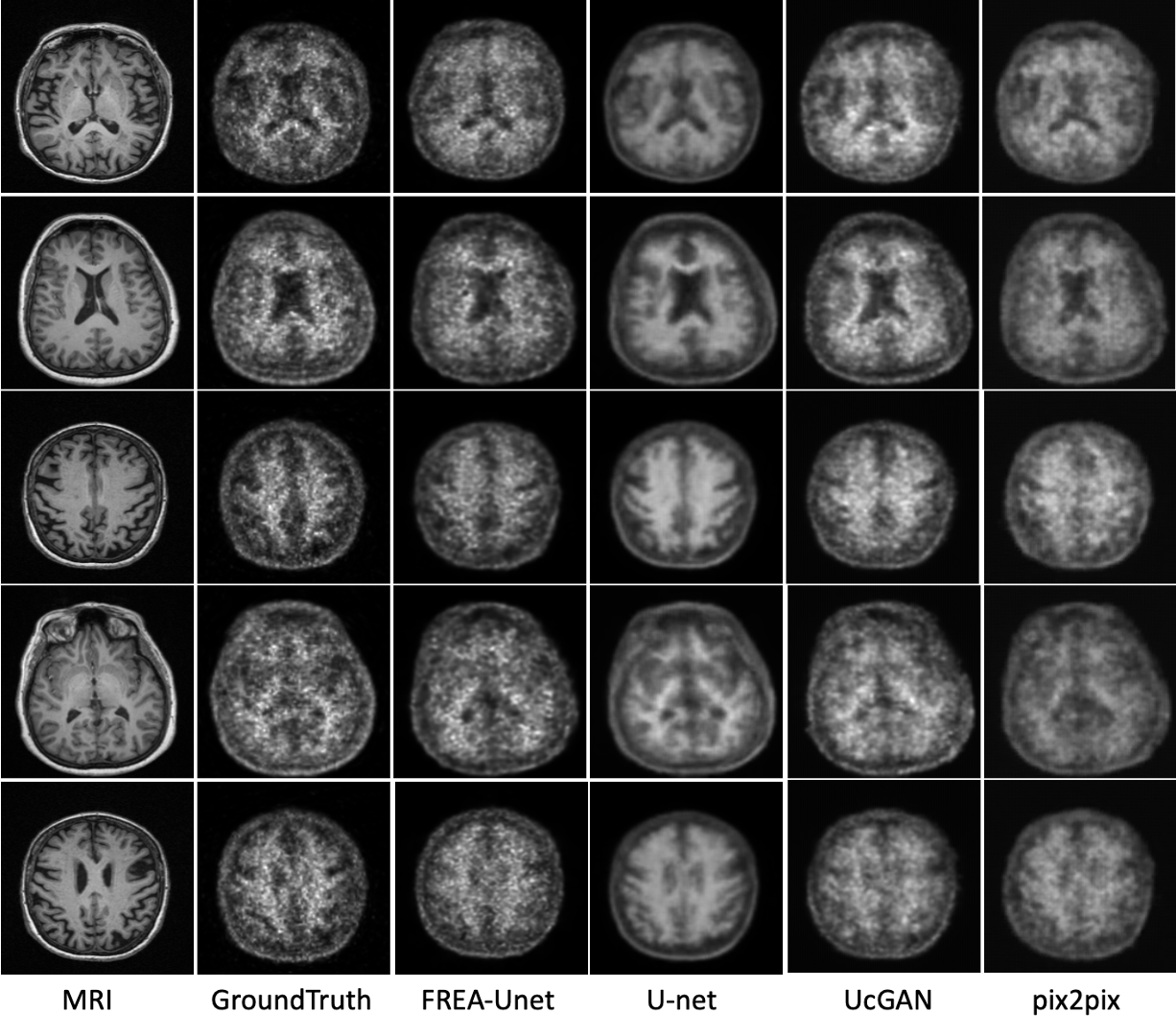} 
\caption{
Qualitative comparison of synthetic PETs generated with FREA-Unet and state-of-the-art models for different patients. The input MRI, corresponding real PET, synPET generated by FREA-Unet, U-net, UcGAN and pix2pix, are shown in the columns, respectively for different patients.}
\label{fig:comparison}
\end{figure}

Fig.~\ref{fig:comparison}, shows generated synPET results using different models for different test cases. The images in the columns are the input MRI, real PET, and the generated PETs using the proposed FREA-Unet model, U-net \cite{han2017mr}, UcGAN \cite{ronneberger2015u}, and pix2pix \cite{isola2017image} respectively. 

\setlength{\tabcolsep}{6pt}
\begin{table}[!h]
\begin{center}
\caption{performance comparison of the FREA-Unet with other models for synthetic PET generation. For each model the average MAEs are computed from entire head. The average PSNR and SSIM are also reported in the second and third rows, respectively.}
\label{table:comparison}
\begin{tabular}{lclclclcl}
\hline
Method $\qquad\qquad$& MAE & PSNR & SSIM\\
\hline
U-net & 52.56 $\pm$ 6.71 & 26.75 $\pm$ 1.96 & 0.84 $\pm$ 0.04\\
pix2pix & 59.19 $\pm$ 7.89 & 24.86 $\pm$ 2.34 & 0.79 $\pm$ 0.04\\
UcGAN & 55.31 $\pm$ 6.25 & 26.11 $\pm$ 2.21 & 0.82 $\pm$ 0.05\\
FREA-Unet & {\bf 46.26 $\pm$ 5.03} & {\bf 28.45 $\pm$ 2.01} & {\bf 0.87 $\pm$ 0.04} \\
\hline
\end{tabular}
\end{center}
\end{table}
\setlength{\tabcolsep}{1.6pt}

Qualitative comparisons for different methods show that the synthetic PETs generated by the proposed FREA-Unet are more accurate in estimating anatomical structures and have more preserved details compared to synPETs generated by other models. Since FREA-Unet generates low/high frequency scales separately, its generated results have higher spatial resolution, and have a successful translation in generating sharp regions including boundaries as shown in Fig.~\ref{fig:comparison}.

The average MAE, PSNR, and SSIM metrics computed based on the real and synthetic PETs for all test cases are listed in Table 2 for each of the aforementioned methods.
The FREA-Unet achieved the best performance in synthetic PET generation compared to other models with the average MAE of 46.26$\pm$5.03 for all test cases.

\subsection{Conclusions}
In this paper, we propose a frequency-aware U-net model, FREA-Unet, for modality transformation in medical imaging. 
FREA-Unet optimizes on different scales separately and takes advantage of attention mechanism to weight different features in low/high scale generation.
Separately optimizing low/high scales using different weights obtained from attention modules helped to generate more accurate synthetic PETs with
preserved organ structures. We have applied this model to estimate PET images from their corresponding MR images on ADNI dataset where we are able to generate sharp synthetic PET images similar to real PET images. 
Our experiments show that the proposed FREA-Unet model can outperform other state-of-the-art methods in modality transfer of medical imaging. 

\bibliographystyle{splncs}
\bibliography{egbib}

\begin{thebibliography}{10}

\bibitem{kang2015prediction}
Kang, J., Gao, Y., Shi, F., Lalush, D.S., Lin, W., Shen, D.:
\newblock Prediction of standard-dose brain pet image by using mri and low-dose
  brain [18f] fdg pet images.
\newblock Medical physics \textbf{42} (2015)  5301--5309

\bibitem{huynh2015estimating}
Huynh, T., Gao, Y., Kang, J., Wang, L., Zhang, P., Lian, J., Shen, D.:
\newblock Estimating ct image from mri data using structured random forest and
  auto-context model.
\newblock IEEE transactions on medical imaging \textbf{35} (2015)  174--183

\bibitem{choi2018stargan}
Choi, Y., Choi, M., Kim, M., Ha, J.W., Kim, S., Choo, J.:
\newblock Stargan: Unified generative adversarial networks for multi-domain
  image-to-image translation.
\newblock In: Proceedings of the IEEE conference on computer vision and pattern
  recognition. (2018)  8789--8797

\bibitem{huang2018multimodal}
Huang, X., Liu, M.Y., Belongie, S., Kautz, J.:
\newblock Multimodal unsupervised image-to-image translation.
\newblock In: Proceedings of the European Conference on Computer Vision (ECCV).
  (2018)  172--189

\bibitem{isola2017image}
Isola, P., Zhu, J.Y., Zhou, T., Efros, A.A.:
\newblock Image-to-image translation with conditional adversarial networks.
\newblock In: Proceedings of the IEEE conference on computer vision and pattern
  recognition. (2017)  1125--1134

\bibitem{liu2017unsupervised}
Liu, M.Y., Breuel, T., Kautz, J.:
\newblock Unsupervised image-to-image translation networks.
\newblock In: Advances in neural information processing systems. (2017)
  700--708

\bibitem{murez2018image}
Murez, Z., Kolouri, S., Kriegman, D., Ramamoorthi, R., Kim, K.:
\newblock Image to image translation for domain adaptation.
\newblock In: Proceedings of the IEEE Conference on Computer Vision and Pattern
  Recognition. (2018)  4500--4509

\bibitem{zhu2017unpaired}
Zhu, J.Y., Park, T., Isola, P., Efros, A.A.:
\newblock Unpaired image-to-image translation using cycle-consistent
  adversarial networks.
\newblock In: Proceedings of the IEEE international conference on computer
  vision. (2017)  2223--2232

\bibitem{zhu2017toward}
Zhu, J.Y., Zhang, R., Pathak, D., Darrell, T., Efros, A.A., Wang, O.,
  Shechtman, E.:
\newblock Toward multimodal image-to-image translation.
\newblock In: Advances in neural information processing systems. (2017)
  465--476

\bibitem{nie2016estimating}
Nie, D., Cao, X., Gao, Y., Wang, L., Shen, D.:
\newblock Estimating ct image from mri data using 3d fully convolutional
  networks.
\newblock In: Deep Learning and Data Labeling for Medical Applications.
\newblock Springer (2016)  170--178

\bibitem{han2017mr}
Han, X.:
\newblock Mr-based synthetic ct generation using a deep convolutional neural
  network method.
\newblock Medical physics \textbf{44} (2017)  1408--1419

\bibitem{emami2018generating}
Emami, H., Dong, M., Nejad-Davarani, S.P., Glide-Hurst, C.K.:
\newblock Generating synthetic cts from magnetic resonance images using
  generative adversarial networks.
\newblock Medical physics \textbf{45} (2018)  3627--3636

\bibitem{emami2020attentionArxive}
Emami, H., Dong, M., Glide-Hurst, C.K.:
\newblock Attention-guided generative adversarial network to address atypical
  anatomy in modality transfer.
\newblock arXiv preprint arXiv:2006.15264 (2020)

\bibitem{li2014deep}
Li, R., Zhang, W., Suk, H.I., Wang, L., Li, J., Shen, D., Ji, S.:
\newblock Deep learning based imaging data completion for improved brain
  disease diagnosis.
\newblock In: International Conference on Medical Image Computing and
  Computer-Assisted Intervention, Springer (2014)  305--312

\bibitem{pan2018synthesizing}
Pan, Y., Liu, M., Lian, C., Zhou, T., Xia, Y., Shen, D.:
\newblock Synthesizing missing pet from mri with cycle-consistent generative
  adversarial networks for alzheimer’s disease diagnosis.
\newblock In: International Conference on Medical Image Computing and
  Computer-Assisted Intervention, Springer (2018)  455--463

\bibitem{armanious2020medgan}
Armanious, K., Jiang, C., Fischer, M., K{\"u}stner, T., Hepp, T., Nikolaou, K.,
  Gatidis, S., Yang, B.:
\newblock Medgan: Medical image translation using gans.
\newblock Computerized Medical Imaging and Graphics \textbf{79} (2020)  101684

\bibitem{ronneberger2015u}
Ronneberger, O., Fischer, P., Brox, T.:
\newblock U-net: Convolutional networks for biomedical image segmentation.
\newblock In: International Conference on Medical image computing and
  computer-assisted intervention, Springer (2015)  234--241

\bibitem{goodfellow2014generative}
Goodfellow, I., Pouget-Abadie, J., Mirza, M., Xu, B., Warde-Farley, D., Ozair,
  S., Courville, A., Bengio, Y.:
\newblock Generative adversarial nets.
\newblock In: Advances in neural information processing systems. (2014)
  2672--2680

\bibitem{gehari2018generating}
Gehari, H.E., Nejad-Davarani, S., Dong, M., Glide-Hurst, C.:
\newblock Generating synthetic cts from magnetic resonance images using
  generative adversarial networks.
\newblock In: Medical physics. Volume~45., WILEY 111 RIVER ST, HOBOKEN
  07030-5774, NJ USA (2018)  E131--E131

\bibitem{nie2017medical}
Nie, D., Trullo, R., Lian, J., Petitjean, C., Ruan, S., Wang, Q., Shen, D.:
\newblock Medical image synthesis with context-aware generative adversarial
  networks.
\newblock In: International Conference on Medical Image Computing and
  Computer-Assisted Intervention, Springer (2017)  417--425

\bibitem{emami2020attention}
Emami, H., Dong, M., Glide-Hurst, C.K.:
\newblock Attention-guided generative adversarial network to address atypical
  anatomy in synthetic ct generation.
\newblock In: 2020 IEEE 21st International Conference on Information Reuse and
  Integration for Data Science (IRI), IEEE (2020)  188--193

\bibitem{wolterink2017deep}
Wolterink, J.M., Dinkla, A.M., Savenije, M.H., Seevinck, P.R., van~den Berg,
  C.A., I{\v{s}}gum, I.:
\newblock Deep mr to ct synthesis using unpaired data.
\newblock In: International workshop on simulation and synthesis in medical
  imaging, Springer (2017)  14--23

\bibitem{rensink2000dynamic}
Rensink, R.A.:
\newblock The dynamic representation of scenes.
\newblock Visual cognition \textbf{7} (2000)  17--42

\bibitem{xiao2015application}
Xiao, T., Xu, Y., Yang, K., Zhang, J., Peng, Y., Zhang, Z.:
\newblock The application of two-level attention models in deep convolutional
  neural network for fine-grained image classification.
\newblock In: Proceedings of the IEEE conference on computer vision and pattern
  recognition. (2015)  842--850

\bibitem{zhou2016learning}
Zhou, B., Khosla, A., Lapedriza, A., Oliva, A., Torralba, A.:
\newblock Learning deep features for discriminative localization.
\newblock In: Proceedings of the IEEE conference on computer vision and pattern
  recognition. (2016)  2921--2929

\bibitem{chen2016attention}
Chen, L.C., Yang, Y., Wang, J., Xu, W., Yuille, A.L.:
\newblock Attention to scale: Scale-aware semantic image segmentation.
\newblock In: Proceedings of the IEEE conference on computer vision and pattern
  recognition. (2016)  3640--3649

\bibitem{mejjati2018unsupervised}
Mejjati, Y.A., Richardt, C., Tompkin, J., Cosker, D., Kim, K.I.:
\newblock Unsupervised attention-guided image-to-image translation.
\newblock In: Advances in Neural Information Processing Systems. (2018)
  3693--3703

\bibitem{chen2018attention}
Chen, X., Xu, C., Yang, X., Tao, D.:
\newblock Attention-gan for object transfiguration in wild images.
\newblock In: Proceedings of the European Conference on Computer Vision (ECCV).
  (2018)  164--180

\bibitem{emami2020spa}
Emami, H., Aliabadi, M.M., Dong, M., Chinnam, R.:
\newblock Spa-gan: Spatial attention gan for image-to-image translation.
\newblock IEEE Transactions on Multimedia (2020)

\bibitem{eriguchi2016tree}
Eriguchi, A., Hashimoto, K., Tsuruoka, Y.:
\newblock Tree-to-sequence attentional neural machine translation.
\newblock arXiv preprint arXiv:1603.06075 (2016)

\bibitem{lin2017structured}
Lin, Z., Feng, M., Santos, C.N.d., Yu, M., Xiang, B., Zhou, B., Bengio, Y.:
\newblock A structured self-attentive sentence embedding.
\newblock arXiv preprint arXiv:1703.03130 (2017)

\bibitem{siridhipakul2019multi}
Siridhipakul, C., Vateekul, P.:
\newblock Multi-step power consumption forecasting in thailand using dual-stage
  attentional lstm.
\newblock In: 2019 11th International Conference on Information Technology and
  Electrical Engineering (ICITEE), IEEE (2019)  1--6

\bibitem{aliabadi2020attention}
Aliabadi, M.M., Emami, H., Dong, M., Huang, Y.:
\newblock Attention-based recurrent neural network for multistep-ahead
  prediction of process performance.
\newblock Computers \& Chemical Engineering (2020)  106931

\bibitem{simonyan2013deep}
Simonyan, K., Vedaldi, A., Zisserman, A.:
\newblock Deep inside convolutional networks: Visualising image classification
  models and saliency maps.
\newblock arXiv preprint arXiv:1312.6034 (2013)

\bibitem{zagoruyko2016paying}
Zagoruyko, S., Komodakis, N.:
\newblock Paying more attention to attention: Improving the performance of
  convolutional neural networks via attention transfer.
\newblock arXiv preprint arXiv:1612.03928 (2016)

\bibitem{jetley2018learn}
Jetley, S., Lord, N.A., Lee, N., Torr, P.H.:
\newblock Learn to pay attention.
\newblock arXiv preprint arXiv:1804.02391 (2018)

\bibitem{zhang2018self}
Zhang, H., Goodfellow, I., Metaxas, D., Odena, A.:
\newblock Self-attention generative adversarial networks.
\newblock arXiv preprint arXiv:1805.08318 (2018)

\bibitem{stone1974cross}
Stone, M.:
\newblock Cross-validatory choice and assessment of statistical predictions.
\newblock Journal of the Royal Statistical Society: Series B (Methodological)
  \textbf{36} (1974)  111--133

\end{thebibliography}

\end{document}